# Physics Data Management Tools for Monte Carlo Transport: Computational Evolutions and Benchmarks

Mincheol Han, Maria Grazia Pia, Hee Seo, Lorenzo Moneta and Chan Hyeong Kim

*Abstract*–**The development of a package for the management of physics data is described: its design, implementation and computational benchmarks. This package improves the data management tools originally developed for Geant4 physics models based on the EADL, EEDL and EPDL97 data libraries. The implementation exploits recent evolutions of the C++ libraries appearing in the C++0x draft, which are intended for inclusion in the next C++ ISO Standard. The new tools improve the computational performance of physics data management.**

## I. Introduction

Data libraries, consisting of tabulations of physics quantities originating from experimental or theoretical sources, are widely used in Monte Carlo simulation. This paper addresses the issue of physics data management in the context of a large scale Monte Carlo system; it reports the results of R&D (research and development) in this domain aimed at improving the computational performance.

For the purpose of this research a test case was identified in the Geant4 [1][2] toolkit: the management of physics data in one of Geant4 electromagnetic physics packages, which heavily exploits data libraries to model physics interactions. Possible improvements to the software design and implementation were investigated, and the associated computational performance was benchmarked with respect to the current Geant4 implementation.

The evaluation concerned programming techniques, like the use of templates, which were sparingly used in the original Geant4 design [3] due to their limited support by many C++ compilers in the early 90's, and new features, which are foreseen for inclusion in the forthcoming revision of the C++ programming language [4].

Manuscript received November 17, 2010. This work was partly supported by the Nuclear Research and Development Program in Korea through the Radiation Technology Development Program and the Basic Atomic Energy Research Institute (BAERI). Support also was received from the Korean Ministry of Knowledge Economy (2008-P-EP-HM-E-06-0000)/the Sunkwang Atomic Energy Safety Co., Ltd.

M. Han, H. Seo and C. H. Kim are with the Department of Nuclear Engineering, Hanyang University, Seoul 133-791, Korea (e-mail: mchan@hanyang.ac.kr, shee@hanyang.ac.kr; chkim@hanyang.ac.kr).

M. G. Pia is with INFN Sezione di Genova, Via Dodecaneso 33, I-16146 Genova, Italy (telephone: +39 010 3536328, e-mail: MariaGrazia.Pia@ge.infn.it).

L. Moneta is with CERN, CH1211 Geneva 23, Switzerland (e-mail:Lorenzo.Montea@cern.ch).

## II. Data Management Features

### A. Physics Data

The investigation concerns the management of the so-called "Livermore Library" data, which are used in Geant4 low energy electromagnetic physics [5][6] simulation. The data are grouped in three libraries: the Evaluated Electron Data Library (EEDL) [7], the Evaluated Photon Data Library (EPDL97) [8] and the Evaluated Atomic Data Library (EADL) [9]. These data were reformatted for use with Geant4, although keeping a structure which is close to the original data format.

EEDL contains tabulations concerning electron interactions with matter. They include cross sections for Bremsstrahlung and ionization, and distributions to generate the energy spectra of the secondary particles produced by these processes. The original secondary particle spectra have been fitted to analytical expressions for use in Geant4; Geant4 physics data management handles the fit parameters.

EPDL97 contains tabulations concerning photon interactions with matter. They include cross sections for Compton and Rayleigh scattering, photoelectric effect and photon conversion, scattering functions and form factors to model the final state of incoherent and coherent scattering.

EEDL and EPDL97 are structured as tabulations of the relevant data (cross sections, scattering functions, form factors) calculated at fixed energies; the data are provided for each element with atomic number between 1 and 100, and in some cases (e.g. photoelectric and ionization cross sections, ionization product spectra) for each atomic subshell. The library data are used in the course of the simulation execution by interpolating the tabulated values.

EADL contains electron binding energies for atomic subshells, and radiative and non-radiative transition probabilities for the generation of X-ray fluorescence and Auger electron emission. The library data are used directly in the course of the simulation.

### B. Requirements

The design of the software responds to the requirements of data management in the simulation.

Total cross section data are used in the simulation by physics processes to determine the mean free path associated with the corresponding type of interaction. This calculation is performed at each step of a transported particle, for all the processes it can be subject to. For convenience, pre-calculated tabulations are generated in the initialization phase of the simulation, which contain mean free path values at given energies; in this process so-called “macroscopic cross sections”, i.e. cross sections for interaction with a material, are calculated, based on the atomic cross sections originally tabulated in the data libraries.

Subshell cross section data are used in the simulation to determine which atomic subshell is concerned, once the tracking algorithm has identified a process as the interaction actually occurring in a given step.

Scattering functions, form factors and particle energy spectra are used in the generation of the final state associated with the process selected by the tracking algorithm.

Physics data assembled in data library files are loaded in the initialization stage of the simulation; further loading on demand may occur during the simulation execution, should any additional data become necessary (e.g. if new elements are created as part of the simulation set-up). Additional data are pre-calculated, if necessary, based on original tabulations. Data are stored in memory to be available for further use.

Physics data are accessed in the course of the simulation execution as needed by the processes they are associated with for the calculation of the mean free path and the generation of the final state.

### C. *Problem domain analysis and current software design*

The physics data management must provide functionality in response to the requirement of data handling in the simulation. The problem domain analysis led to the configuration of physics data management as a package, distinct from the physics processes using the data. This design solution ensures greater flexibility, since the two domains – physics modeling and data management – can evolve independently. This design supports appropriate usage of design methods, privileging composition over the burden of inheritance to share common data management functionality across different physics objects.

The problem domain decomposition led to the identification of well defined responsibilities, which were associated with objects:

- low level data management, involving basic operations like loading and accessing the data (associated with the *G4VEMDataSet* abstract class and derived classes),
- data interpolation, with optional algorithms (handled by the *G4VInterpolationAlgorithm* abstract class and derived classes,
- manipulation of data to deal with materials (associated with the *G4VCrossSectionHandler* abstract class and derived classes).

The design of the original data management system used in Geant4 is based on the Composite design pattern [10]. This pattern allows the transparent treatment of different type of data: “leaf” data types, associated with the lowest level physics entity pertinent to the related process (e.g. subshell data) and composite data, which in turn contain lower level data (e.g. atomic data, where each atomic data set may involve subshell data sets).

The provision of alternative interpolation algorithms through a transparent mechanism is handled through a Strategy pattern [10]. A Prototype pattern [10] is used to clone the selected interpolation algorithm across composite objects.

This design was originally used in Geant4 low energy electromagnetic package to deal with EEDL, EPDL97 and EADL data; it was then applied also to other physics data [11] in that package, including data associated with reengineering physics models originally implemented in Penelope [12].

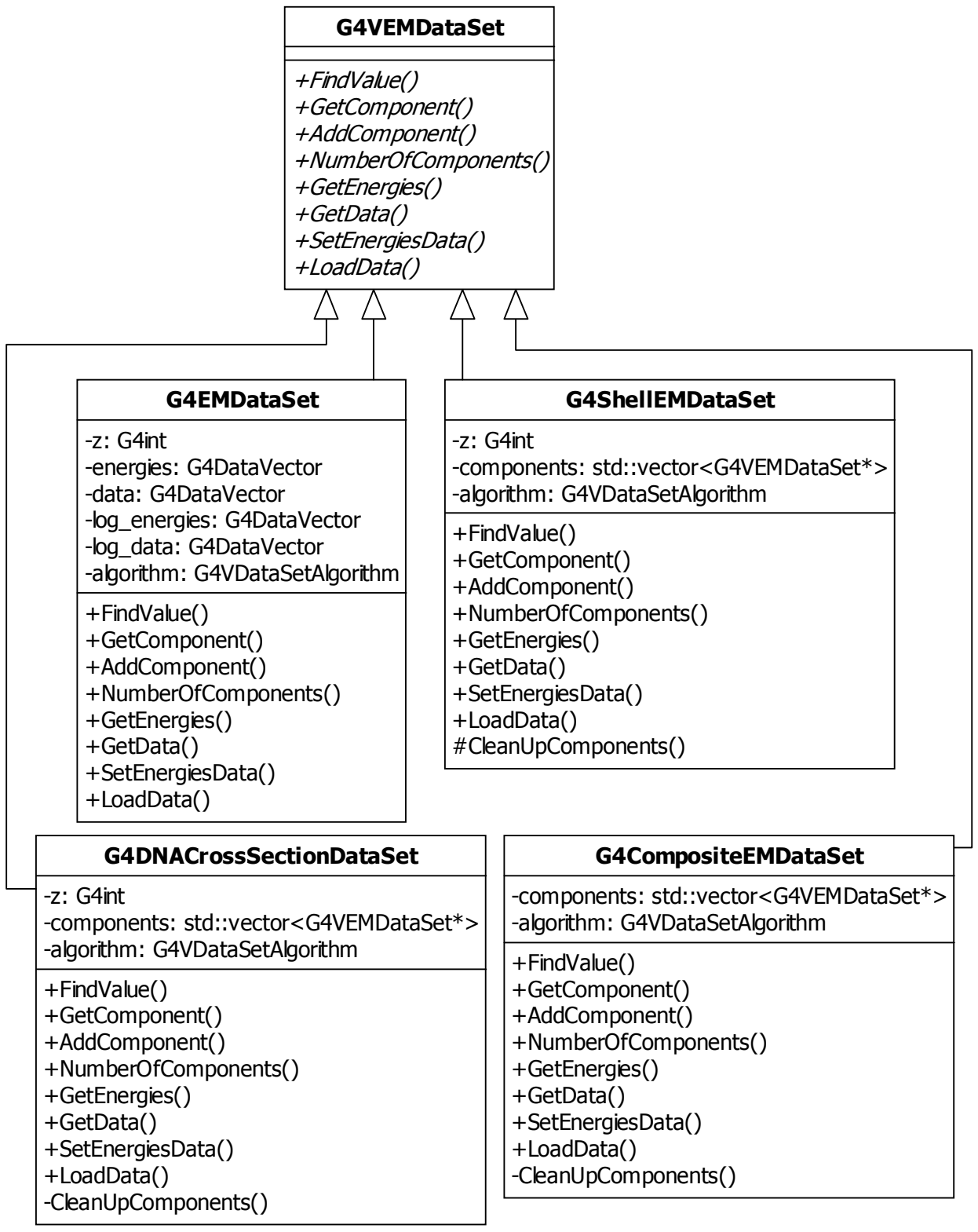

Fig. 1 UML class diagram illustrating basic data management.

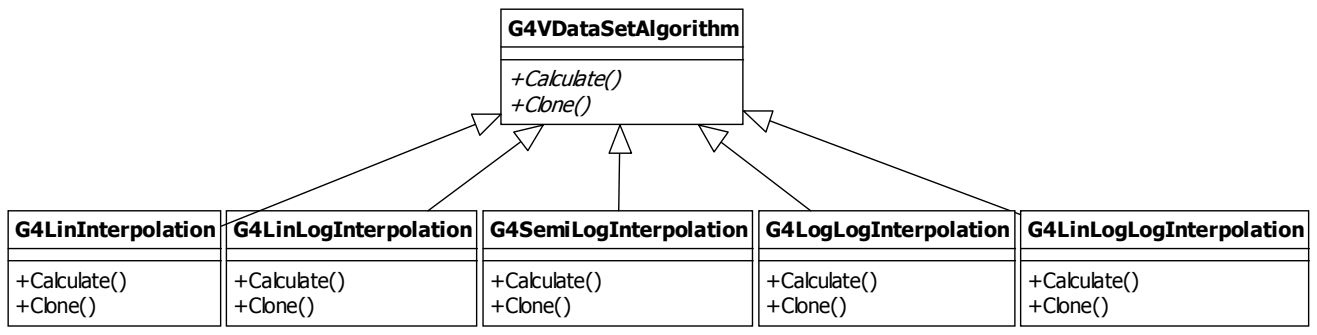

Fig. 2 Class diagram illustrating the management of a variety of interpolation methods.

A class diagram illustrating the main features of the software design is shown in Fig. 1; a class diagram documenting the design of interpolation is shown in Fig. 2; the design features concerning the manipulation of data related with materials are shown in Fig. 3. The diagrams are expressed in UML (Unified Modeling Language) [13].

## III. R&D TOPICS AND BENCHMARKS

Various physics data structures and software design features were investigated to evaluate whether they would contribute to improve the computational performance of the data management software.

While computational performance was the main objective driving the study, the intent of simplifying the software design provided complementary motivation for the R&D described in the following sections. More agile software facilitates its maintenance and possible future evolution; it also supports the transparency of its semantics, thus facilitating its appropriate use.

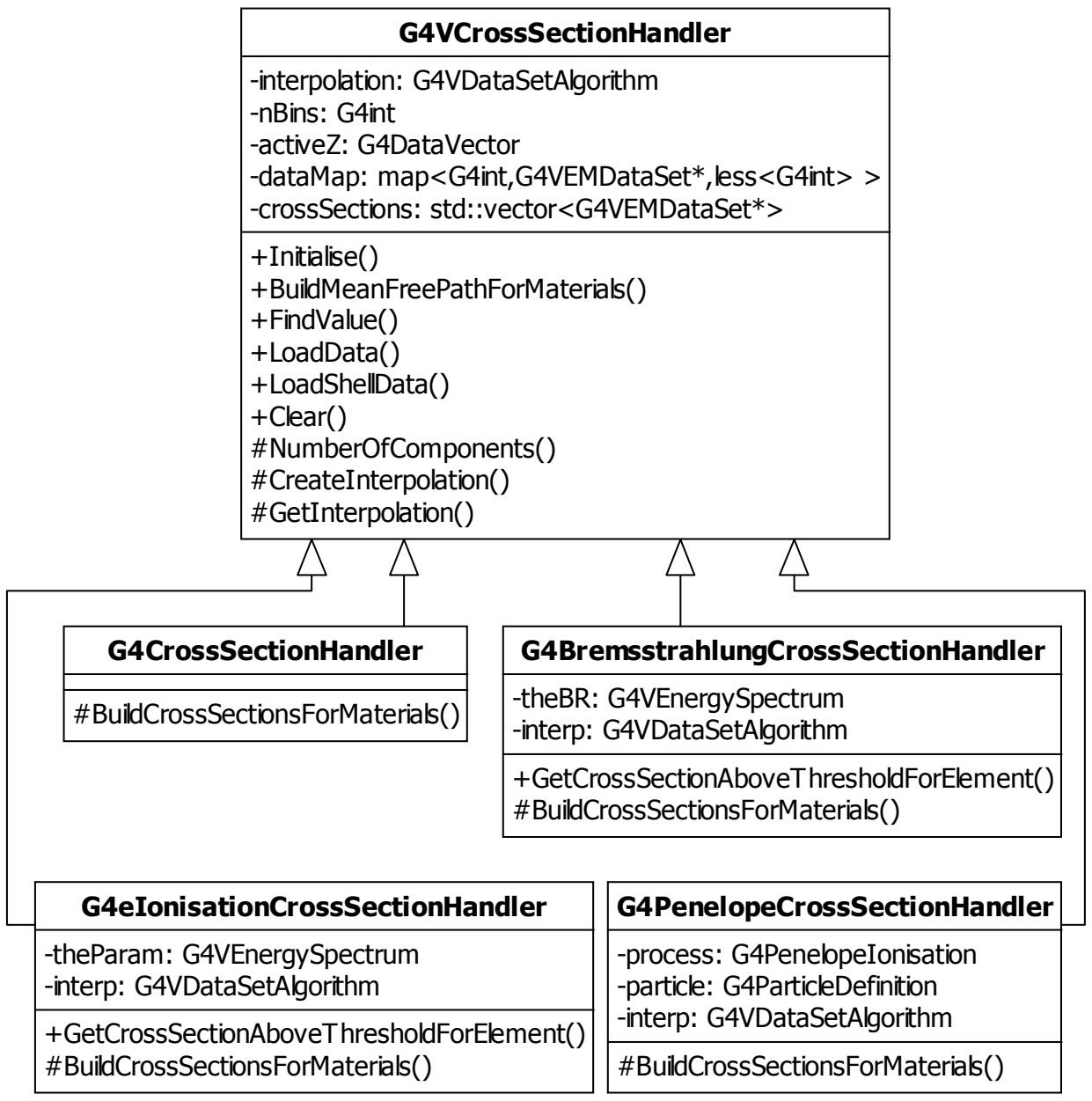

Fig. 3 Class diagram illustrating the management of data dealing with materials.

### *A. Test configuration*

The benchmarks were based on software released in Geant4 9.4.beta version and data released in G4EMLOW6.13.

The full set of benchmarks were executed on a Intel® Core™ Duo CPU E8500 equipped with a 3.16 GHz processor, with 4 GB of memory, running under Linux SLC5. GNU C++ compiler gcc 4.3.5 was used in this configuration. A subset of tests were executed on a Microsoft Windows system configured with Intel® CPU U4100 with 1.30GHz processor, with 1.96GB of memory, running under Windows XP SP3. The MSVC++9 C++ compiler (with SP1) was used for these tests.

Two types of tests were performed to evaluate the computational performance of the code respectively at loading and retrieving data. The "load" test consisted of loading the data corresponding to a number of instantiated elements between 1 and 100; each experiment was repeated 100 times, and the whole series was repeated 10 times. The "retrieve" test consisted of finding the data associated with a randomly chosen atomic number; the finding procedure was repeated one million times, and the whole experiment was repeated 10 times.

### *B. Data structure*

Some of the original data are structured as a single file, which encompasses data for one hundred elements. Loading the data needed for the elements required in a simulation, i.e. corresponding to the materials present in the experimental set-up, requires parsing the whole data file; this input-output (I/O) operation is expensive. If the data are loaded on demand, that is, when the data pertinent to an element become necessary during the simulation execution, the expensive parsing procedure may be repeated several times in the course of a simulation.

To improve the agility of the loading process, such data files were split into individual files, each one associated with one element. An example of the improved data loading performance is shown in Fig. 4.

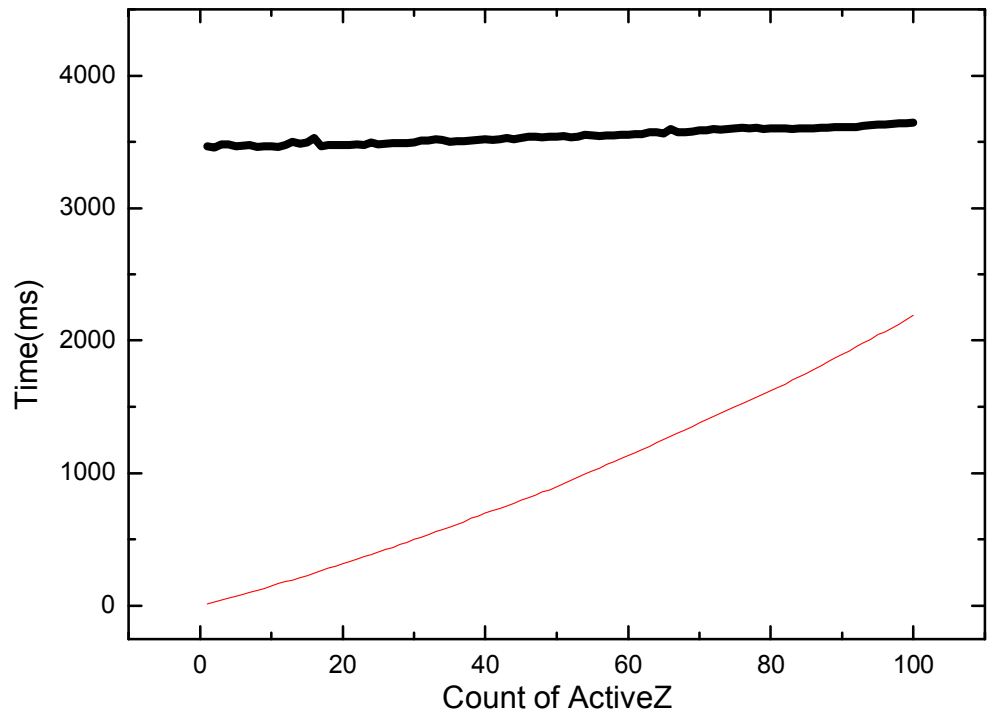

Fig. 4 An example of the optimization of data loading by splitting the excitation data file into individual files containing data for a single element: time to load the original data (thick black histogram) and split data (thin red histogram), as a function of the number of instantiated elements.

The performance of data management is affected by the quantity of physics data to be handled. Large physics tabulations require large memory allocation for storing the data, time to load them into memory and to search trough them.

A study was performed to evaluate on quantitative ground whether the size of the original data libraries could be reduced without affecting the precision of physics calculations. For this purpose a test was developed to verify if the suppression of a given datum in the tabulations would allow the calculation of values of comparable accuracy through interpolation between adjacent data points, over the whole interval between them. The suppression of an original data point was considered tolerable if one could reproduce the same interpolated values as the original data tabulation within 0.01%.

An example of the reduction of the data size and its effect on the computational performance at loading the data is shown in Fig. 5 and Fig. 6.

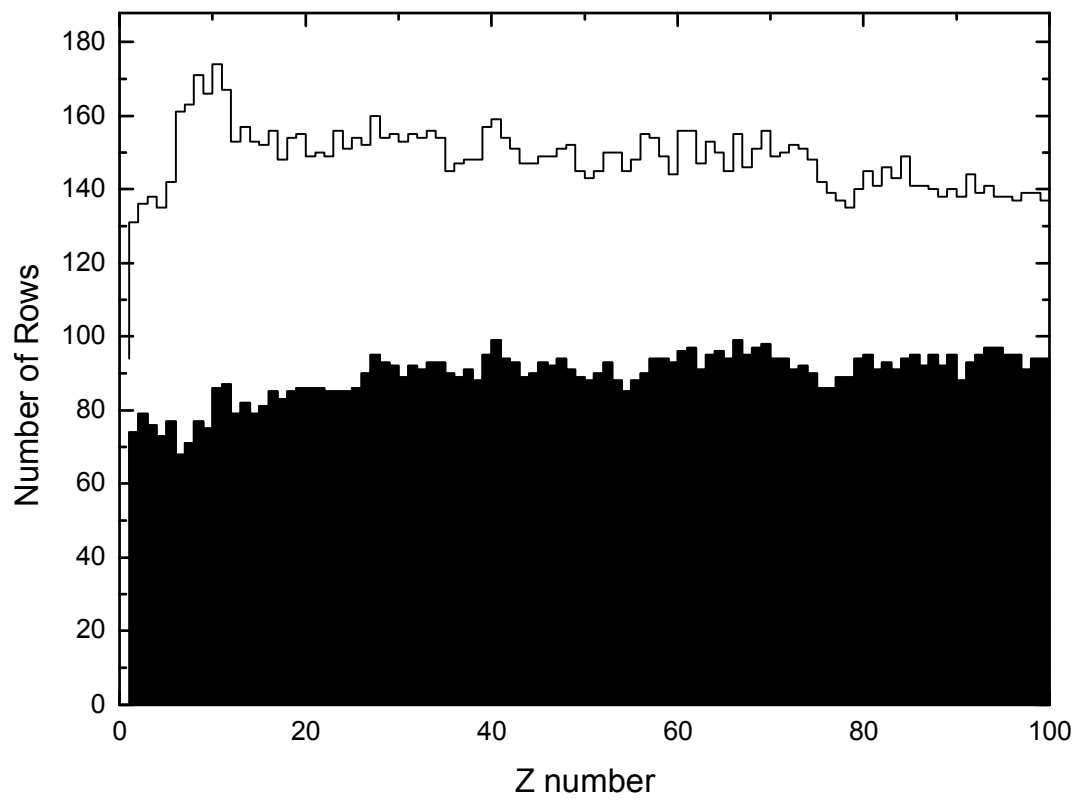

Fig. 5 Optimization of the data library size regarding Compton scattering functions: original number of data (white histogram) and reduced data (black histogram), as a function of the atomic number.

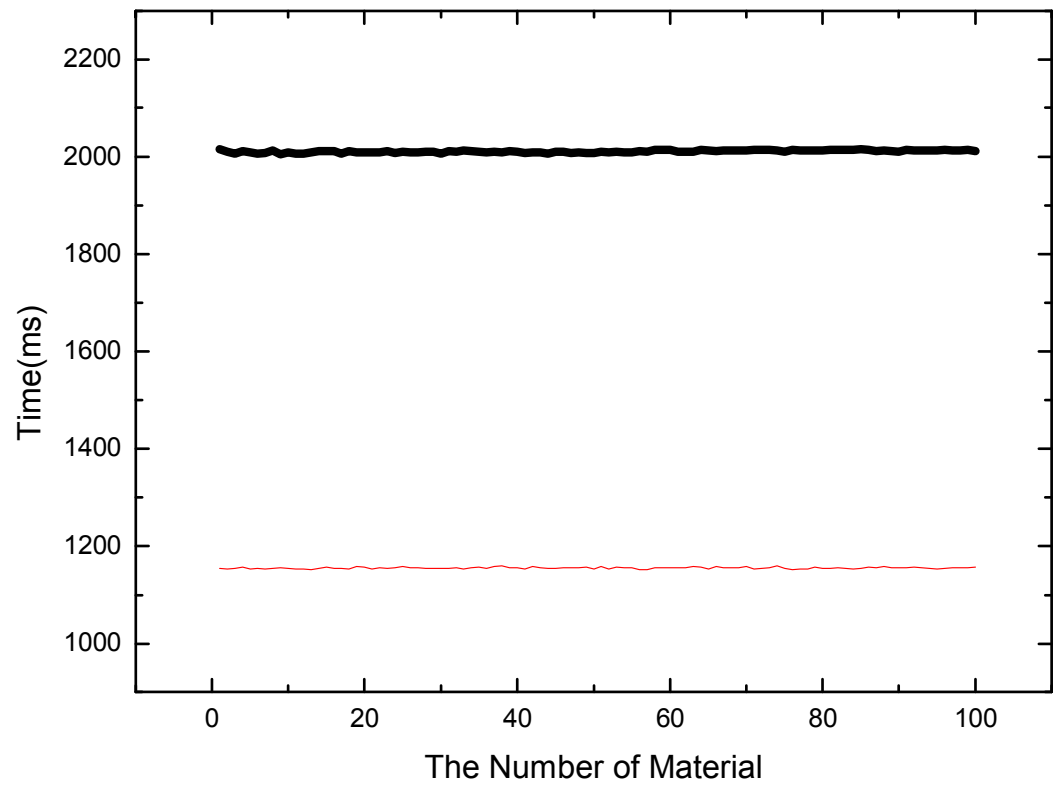

Fig. 6 Optimization of data loading performance due to the reduction of Compton scattering functions data: original number of data (thick black histogram) and reduced data (thin red histogram), as a function of the number of instantiated elements.

## C. *Use of forthcoming C++ features*

The forthcoming edition of the C++ language includes several new features. The R&D project evaluated whether the data management implementation could profit from a type of container, a so-called "hash map", which is not available in the current Standard Template Library (STL), but is foreseen for inclusion in the new C++ standard.

The proposed C++0x TR1 name for a hash table is unordered_map; it will replace the various incompatible implementations of the hash table (called hash_map by the gcc and MSVC compilers). As its name implies, unlike the map class, the elements of an unordered_map are not ordered; this is due to the use of hashing to store objects. The main advantage of hash tables over other types of associative containers is speed.

This container is currently accessible as std::tr1::unordered_map. Until TR1 is officially accepted into the upcoming C++0x standard, unordered_map is available from the <tr1/unordered_map> header file and from <unordered_map> in MSVC. unordered_map can be used in a similar way to the map class in C++ STL.

The map containers currently used in Geant4 data management system were replaced by unordered_map ones. This modification has negligible effects on data loading performance, while it improves significantly the performance of retrieving the data. An example is shown in Fig. 7, which concerns the retrieval of pair production cross section data as a function of the number of instantiated elements.

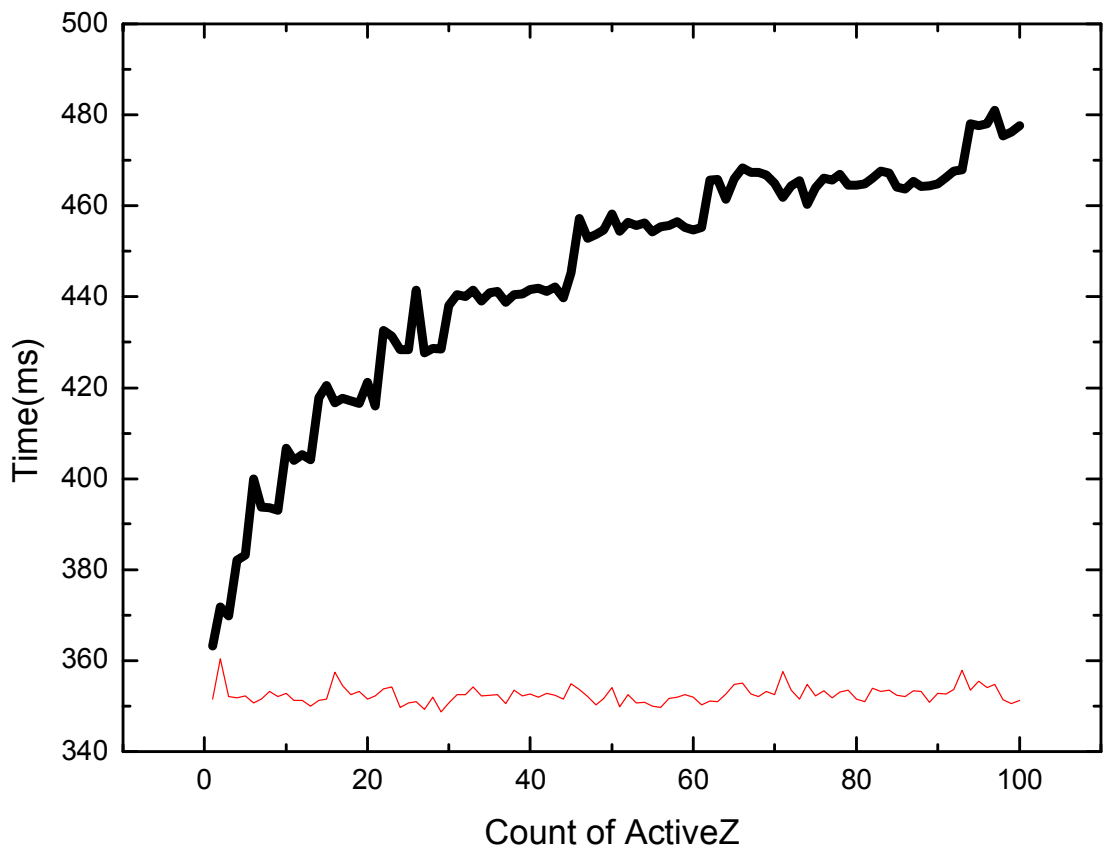

Fig. 7 The retrieval of pair production cross section data as a function of the number of instantiated elements: current Geant4 implementation (thick black curve) using the map STL container and implementation using unordered_map, which is foreseen for inclusion in the forthcoming edition of the C++ language.

## D. *Caching pre-calculated data*

An improvement of the performance of the data management has been announced as part of the Geant4 9.3 release [14]; it consists of caching pre-calculated logarithms in base 10 of the data for use in logarithmic interpolation. The authors of this paper are not responsible for these modifications and do not claim any credit for them.

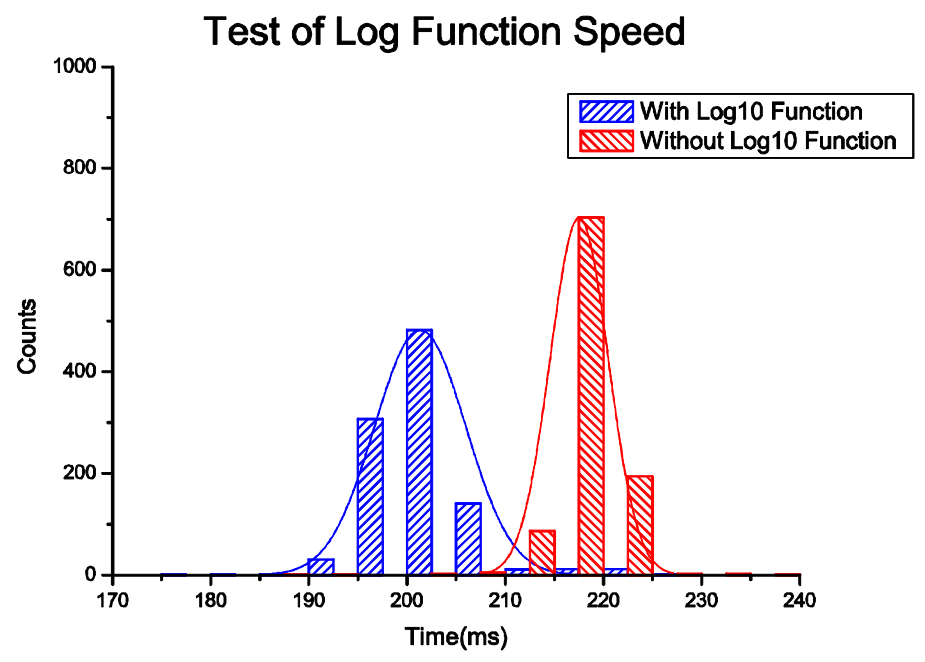

Fig. 8 Computational performance overhead due to the calculation of a single logarithm in base 10.

We performed some tests to evaluate quantitatively the computational performance effects of caching pre-calculated logarithms in base 10 of the data. A test including 100 million calls to a logarithm in base 10 showed that the overhead due to a single call is in average of the order of 10%, as can be seen in Fig. 8. The performance improvement at retrieving data due to caching pre-calculated logarithm in base 10 of the data is shown in Fig. 9; however, caching the data adds some penalty to the data load procedure, as one can see in Fig. 10. Caching additional data involves some penalty in memory usage as well; however, the memory penalty is relatively small compared to the size of a typical Geant4 simulation application (each data chunk corresponds to approximately 100 bytes).

The modifications associated with Geant4 9.3 release introduced a semantic flaw in the design regarding the linear interpolation class; a further flaw was detected in the implementation of the linear-logarithmic class. They were corrected in the software described in this paper.

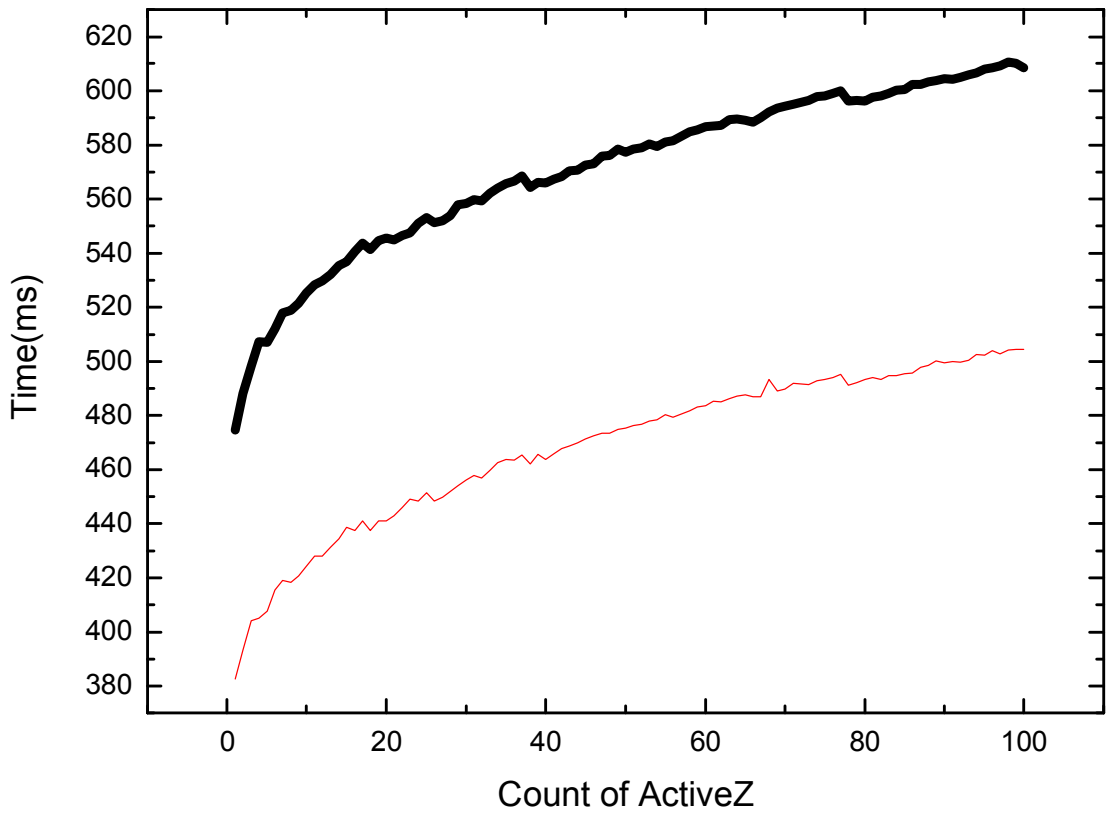

Fig. 9 Performance improvement at retrieving data due to caching pre-calculated logarithms in base 10 of the data for subsequent logarithmic interpolation: with original data (thick black curve) and caching pre-calculate values (thin red curve).

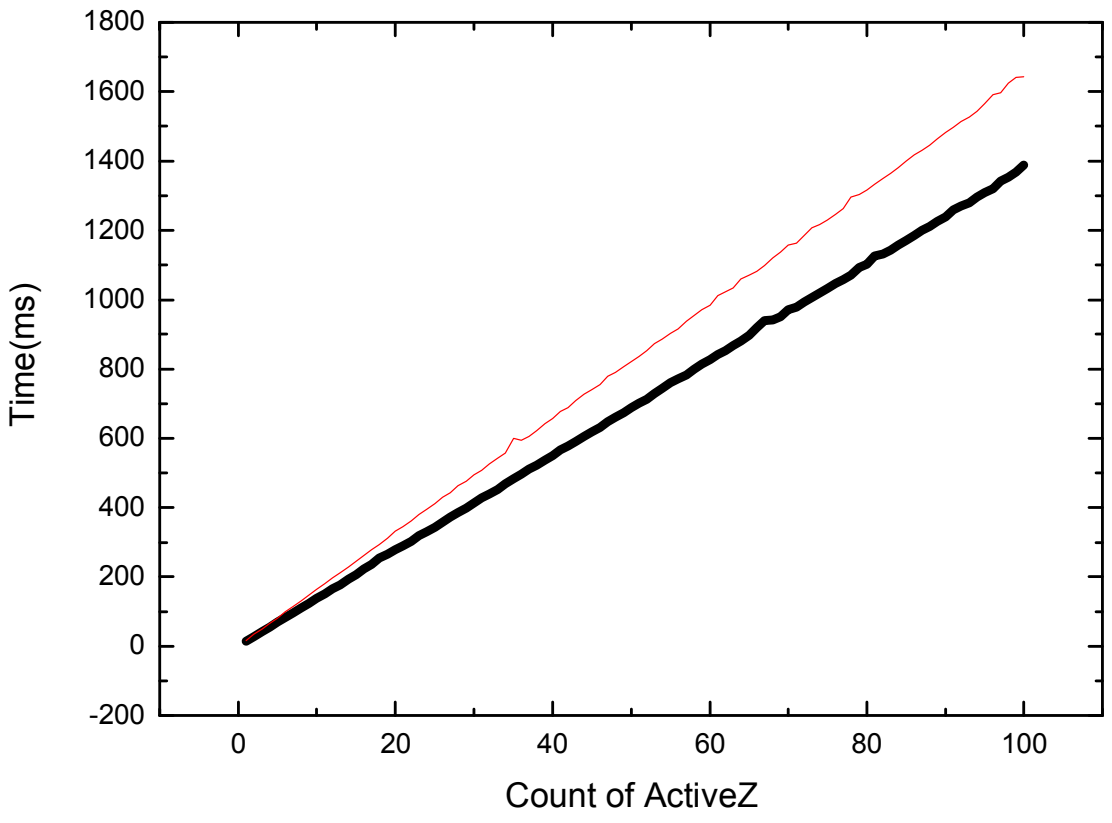

Fig. 10 Performance overhead at loading Compton cross section data due to caching pre-calculated logarithmic values (thin red curve) with respect to loading the original data (thick black curve).

## E. *Software design*

The software design was modified. The handling of data concerning atoms, shells and materials was unified under the responsibility of one class. The problem domain analysis recognized that polymorphic behavior of data sets and interpolation algorithms was not necessary at runtime through dynamic binding, rather could be realized by exploiting template programming techniques. The new design is illustrated in Fig. 11.

The adoption of template programming contributes to improving execution speed, since it eliminates the overhead due to the virtual table associated with inheritance. The performance gain at loading depends on the type of data: in some cases, as in Fig. 12, it is significant, while for some other data types it is negligible.

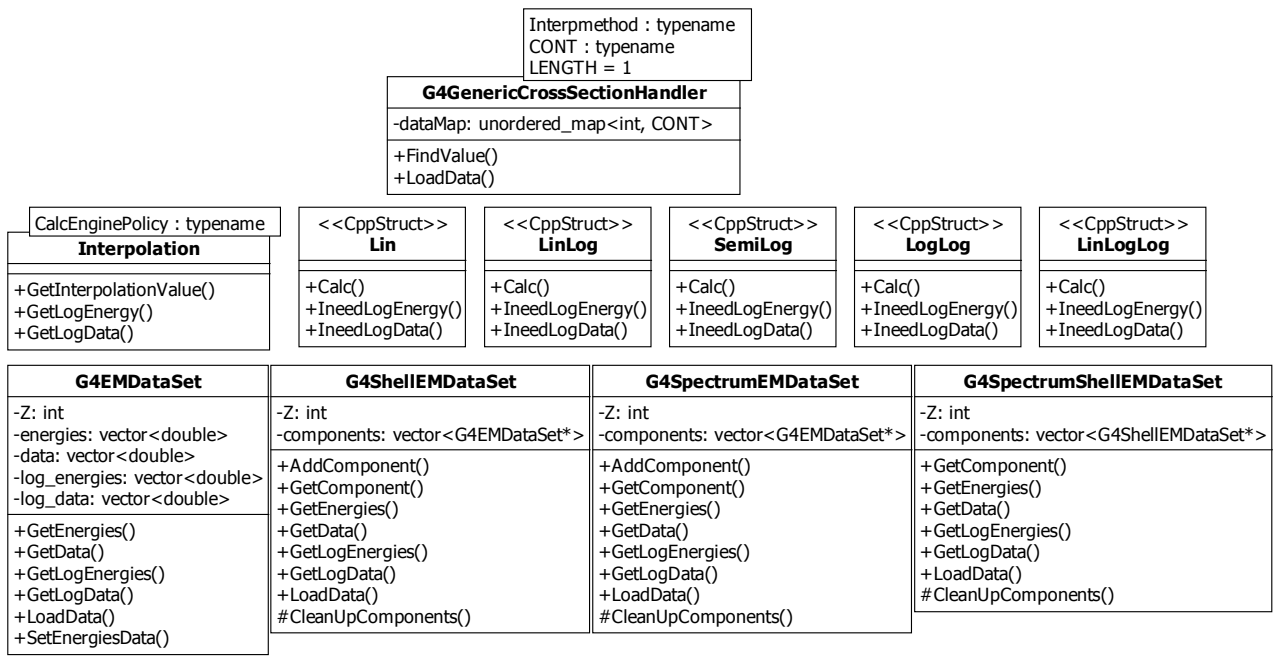

Fig. 11 Improved prototype software design.

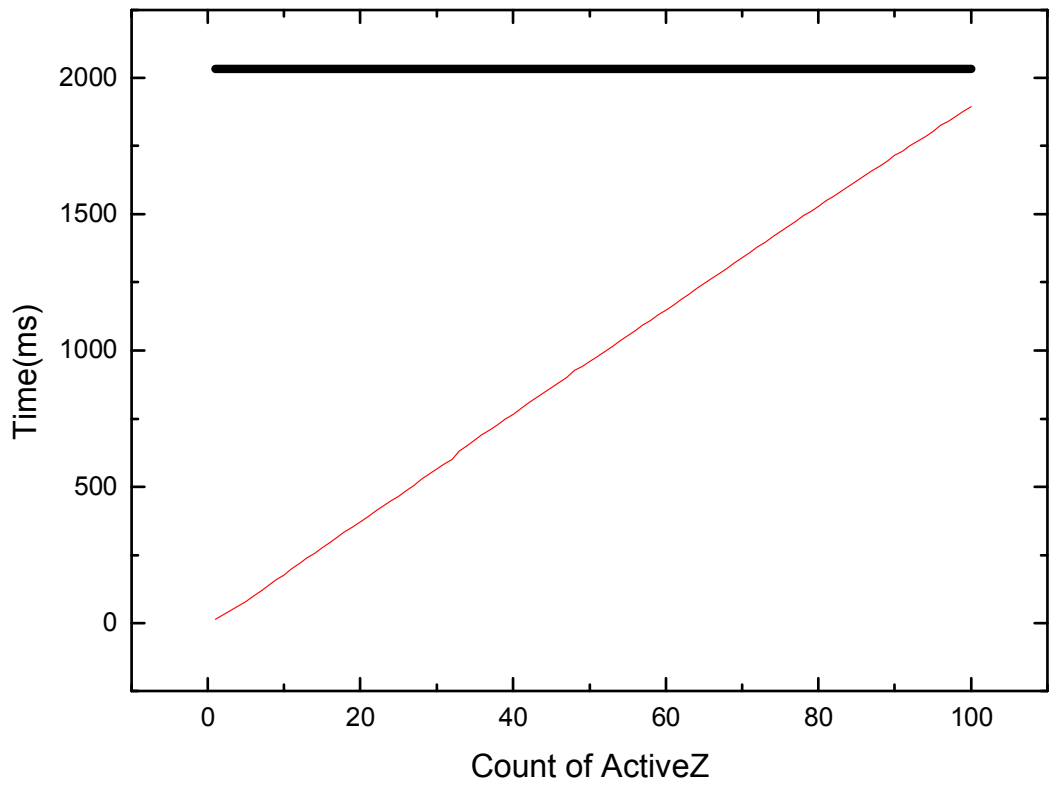

Fig. 12 Performance improvement at loading Compton cross section data (thin red curve) due to the new design using templates, with respect to the original design (thick black curve).

The performance is significantly improved also at data retrieving, as illustrated in Fig. 13 and Fig. 14. These plots also show the combined effect of the new software design and of using the unordered_map foreseen in the new C++ standard.

In some cases, usually corresponding to a relatively small number of data, loading the whole data set corresponding to 100 elements into a STL vector proves to be more effective than loading selected data into an unordered_map, as seen in Fig. 15. Further investigations are in progress.

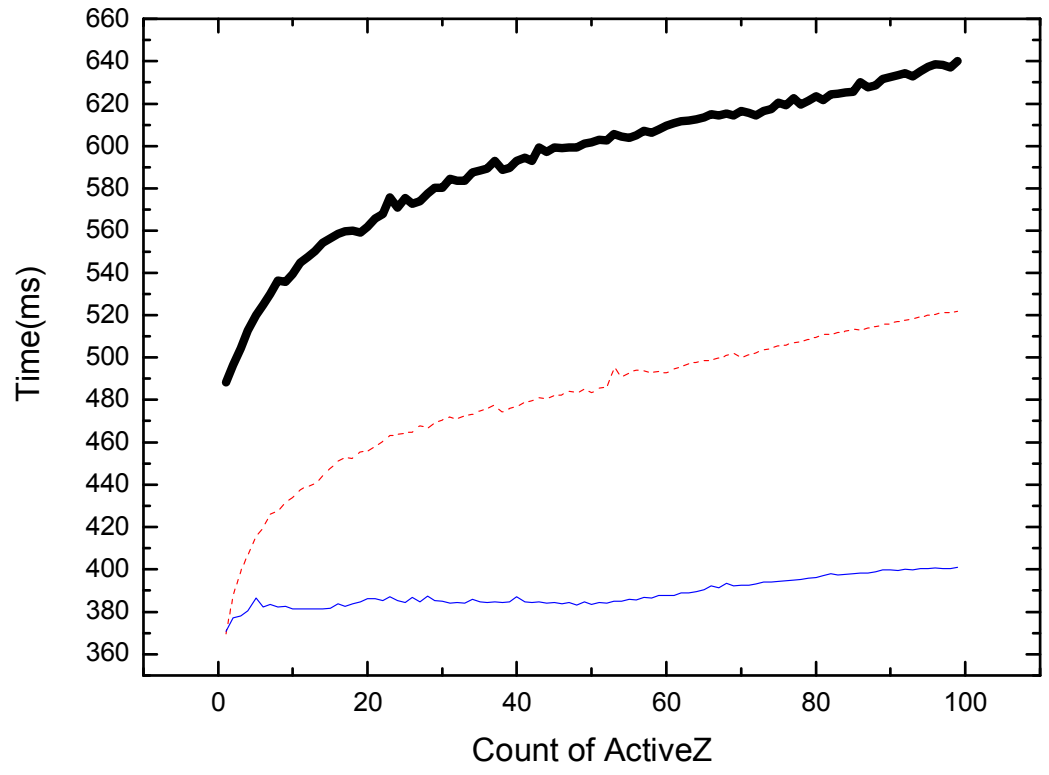

Fig. 13 Improvement at retrieving pair production cross data due to the design based on templates (dashed red curve) and on the use of unordered_map along with the new design (thin blue curve), with respect to the original classes (thick black curve).

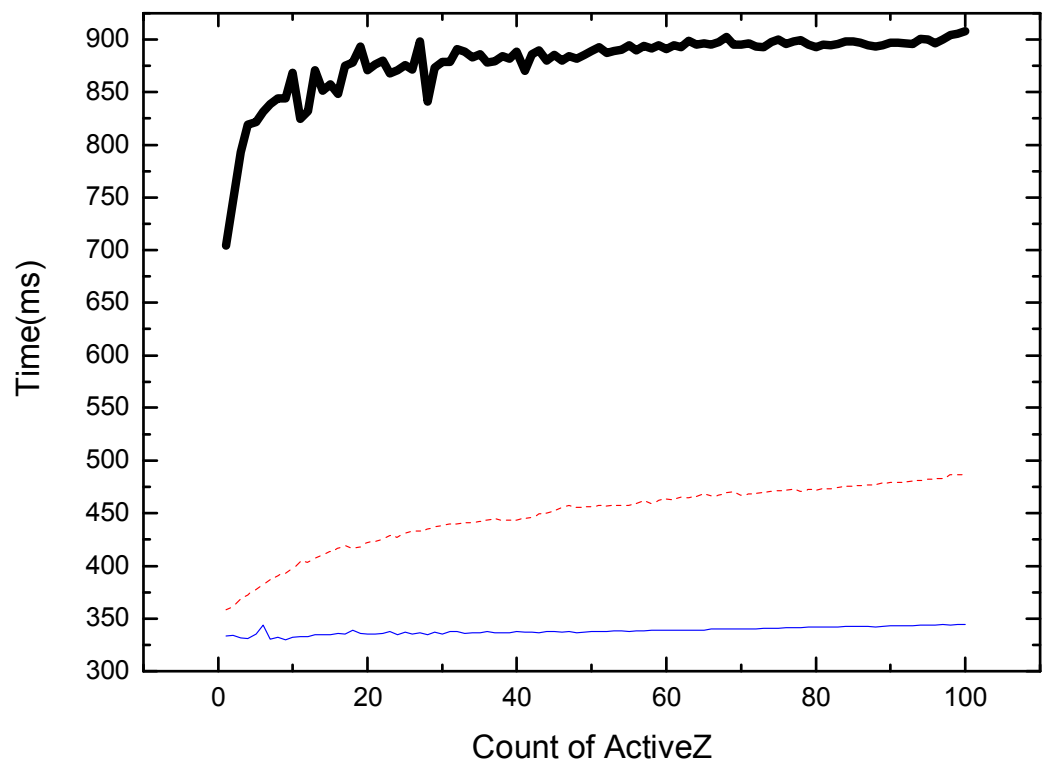

Fig. 14 Improvement at retrieving Bremsstrahlung spectrum data due to the design based on templates (dashed red curve) and on the use of unordered_map along with the new design (thin blue curve), with respect to the original classes (thick black curve).

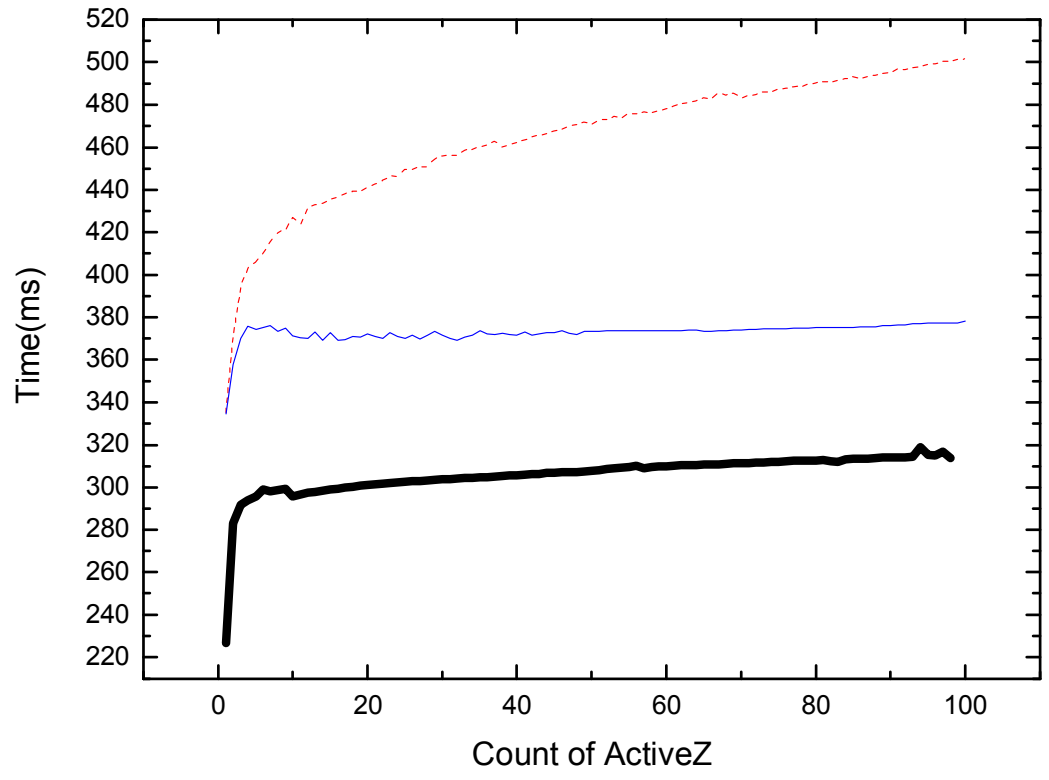

Fig. 15 Performance overhead due to using unordered_map (thin blue line) or a map (red line) in the new design scheme, with respect to the original design using a STL vector and loading all 100 data sets of Rayleigh form factors (thick black line).

## IV. Conclusion

An R&D project evaluated the possibility of improving the design and computational performance of Geant4 physics data management system. Various issues were addressed: the improvement of the structure of the data themselves, the use of new features in the forthcoming C++ standard, the use of template programming to replace the current design based on the Composite design pattern. A recent improvement implemented in Geant4, consisting of caching some pre-calculated data, was quantitatively evaluated.

Tests were performed on a sample corresponding to two data libraries, EEDL and EPDL97, used by Geant4 electromagnetic physics processes.

The effects on computational performance due to all the previously mentioned topics were independently measured. The overall improvement in computational performance is significant, both at loading and retrieving data; depending on the type of data, gains in data retrieving of approximately 30% up to almost a factor 3 can be achieved.

The project described in this paper has achieved significant improvement in the computational performance of physics data management. Key issues for the improvement are a new software design and the use of a container not present in the current STL, but available in the coming new C++ standard. Further benchmarks are foreseen in concrete Geant4 application test cases.

The results of this project suggest that performance improvements could be achieved in other parts of Geant4 through improvements to the software design. New features available in the forthcoming C++ standard could provide opportunities for further improvement.

## Acknowledgment

We thank CERN Directorate for support to the research described in this paper.

## References

[1] S. Agostinelli et al., "Geant4 - a simulation toolkit", *Nucl. Instrum. Meth. A,* vol. 506, no. 3, pp. 250-303, 2003.
[2] J. Allison et al., "Geant4 Developments and Applications", *IEEE Trans. Nucl. Sci.,* vol. 53, no. 1, pp. 270-278, 2006.
[3] S. Giani, GEANT4 : an object-oriented toolkit for simulation in HEP, CERN Report CERN-LHCC-98-044, Geneva, 1998.
[4] C++, ISO/IEC JTC1 SC22 WG21 N3092, Geneva, 2010.
[5] S. Chauvie et al., "Geant4 Low Energy Electromagnetic Physics", Proc. Computing in High Energy and Nuclear Physics, Beijing, pp. 337-340, 2001.
[6] S. Chauvie et al., "Geant4 Low Energy Electromagnetic Physics", Conf. Rec. IEEE Nucl. Sci. Symp., pp. 1881-1885, 2004.
[7] S. T. Perkins et al., "Tables and Graphs of Electron Interaction Cross Sections from 10 eV to 100 GeV derived from the LLNL Evaluated Electron Data Library", UCRL-50400 Vol. 31, 1991.
[8] D. Cullen et al., "EPDL97, the Evaluated Photon Data Library, 97 version", UCRL–50400, Vol. 6, Rev. 5, 1997.
[9] S. T. Perkins et al., "Tables and Graphs of Atomic Subshell and Relaxation Data Derived from the LLNL Evaluated Atomic Data Library (EADL), Z=1-100", UCRL-50400 Vol. 30, 1997.
[10] E. Gamma, R. Helm, R. Johnson, J. Vlissides, Design Patterns, Addison-Wesley, New York, 1995.

[11] S. Chauvie et al., "Geant4 physics processes for microdosimetry simulation: design foundation and implementation of the first set of models", *IEEE Trans. Nucl. Sci.,* vol. 54, no. 6, pp. 2619 - 2628, 2007.
[12] J. Baro, J. Sempau, J. M. Fernandez-Varea and F. Salvat, "Penelope, an algorithm for Monte Carlo simulation of the penetration and energy loss of electrons and positrons in matter", *Nucl. Instrum. Meth. B*, vol. 100, no. 1, pp. 31-46, 1995.
[13] G. Booch, J. Rumbaugh and I. Jacobson, The Unified Modeling Language User Guide, Addison-Wesley, Boston, 1999.
[14] Online: http://cern.ch/geant4 /support/ReleaseNotes4.9.3.html